**Excitation functions and isotopic effects in (n,p) reactions for stable iron isotopes from reaction threshold to 20 MeV.**

J. Joseph Jeremiah [a], Damewan Suchiang [b], B.M. Jyrwa [a,*]

[a] *Department of Physics, North Eastern Hill University, Shillong-793022, Meghalaya, India*

[b] *Department of Physics, Tura Government College, Tura-794001, Meghalaya, India.*

* Corresponding Author. *Email address* : bjyrwa90@hotmail.com (B.M. Jyrwa)

**Abstract**

The excitation functions for (n,p) reactions from reaction threshold to 20 MeV on four stable Iron isotopes viz; $^{54}$Fe,$^{56}$Fe,$^{57}$Fe and $^{58}$Fe were calculated using Talys-1.2 nuclear model code, this essentially involves fitting a set of global parameters. An excellent agreement between the calculated and experimental data is obtained with minimal effort on parameter fitting and only one free parameter called 'Shell damping factor' has been adjusted. This is very importance to the validation of nuclear model approaches with increased predictive power. The systematic decrease in (n, p) cross-sections with increasing neutron number in reactions induced by neutrons on isotopes of iron is explained in terms of the pre-equilibrium model. The compound nucleus and pre-equilibrium reaction mechanism as well as the isotopic effects were also studied extensively.

**Keywords**

Excitation functions, Shell damping factor, Compound nucleus model,

Pre-equilibrium mechanism.

**1. Introduction**

The common materials suitable for the reactor structures are stainless steel with Cr, Fe and Ni as main constituents. Structural materials are used to make the core and several other essential parts of a nuclear reactor. Thus, they must possess a high resistance to mechanical stress, stability to radiation and high temperature, and low neutron absorption. Materials used for structural purposes are Al, Be, C (graphite), Cr, Fe, Mg, Ni, V and Zr. Stainless steels, alloys of Fe, Cr, Ni, and other metals, are by far the most used structural materials since the beginning of the nuclear industry. Stainless steels, Streicher (1979) are the main off-core structural materials for thermal reactors and are essential materials for fast neutron reactors. C and ferritic steels are sued for nuclear pressure vessels in LWR and heavy water reactors (HWR). Lightly alloyed steels are used in PWR reactors. In nuclear fusion power plants, the first wall blanket structure is critical for safety and environmental considerations. The performance of the first wall blanket is dependent on structural material properties. Current designs of these power plants involve steels as the first wall structural materials and assume operating temperatures below 550 $^{\circ}$C. However, a considerable increase in efficiency can be gained by raising this temperature using ferritic-martensitic materials. A good candidate is Iron (Fe) that can operate at 950 $^{\circ}$C. The microstructural evolution of a material under irradiation depends on its crystallographic structure. Most metals used in nuclear structural materials have one of three crystallographic structures, two cubic and one hexagonal. Examples of cubic materials are W, Fe, V and ferritic steels, which are body centred cubic (BCC), as well as Cu, austenitic steels and Ni alloys, which are face centred cubic (FCC).

Precise knowledge of the excitation function and isotopic effects in the (n,p) reaction cross-section on Iron isotopes are very important. Measurements of excitation curves for the (n,p) reaction on chains of isotopes have revealed an interplay of the compound nucleus and the pre-equilibrium mechanisms. The shapes and magnitudes of the excitation functions from the reaction thresholds up to 20 MeV are described by model calculations using a consistent parameter set. The main purpose of this work is to check the predictive power of the nuclear model calculations on the excitation functions and to understand the mechanisms of compound nucleus and pre-equilibrium models starting from reaction threshold to 20 MeV as well as to check the shell effects in the (n, p) reactions on four stable Iron isotopes viz. $^{54}$Fe, $^{56}$Fe, $^{57}$Fe and $^{58}$Fe. All experimental data are taken from EXFOR database.

## 2. Nuclear models and calculations

The nuclear reaction model calculations performed include the direct-interaction, pre-equilibrium and compound nucleus contributions. Essentially the Talys-1.2 code has been used with the systematic originating from global phenomenological analyses. It can be used for the incident energy range from 1 KeV – 200 MeV, for target mass numbers between 10 and 239 and treats n, γ, p, d, t, h and α as projectiles and ejectiles. The execution time depends on the calculations involved, but is normally fast only in seconds for this type of work. In the input file, all the choices can be made and many parameters can be adjusted such as the level density parameter, the shell damping factor and the nuclear models. Much information regarding the choices of input parameters, nuclear models, level density parameters are described and explained in detail in the Talys user manual, Koning et al. (2009). Since nuclear model calculations and fits to experiments generally require many adjustable parameters, it is important that these parameters all remain within physically acceptable limits. The Optical model potentials for neutron and proton used in the Talys-1.2 calculation is the global parameterizations of Koning and Delaroche (2003). We have chosen to study the global predictive power that reflects the state of knowledge of nuclear model parameters.

Level density as shown by Margarit Rizea, et al. (2005) is an important factor in the calculation of cross sections. This is due to typical spacing's of the first excited nuclear level spacing in medium and heavy nuclei ranges from a tenth to a few tens of MeV, for low excitation energies, but becomes very densely spaced when their energies increase. This clearly shows that an individual description is not possible, which justifies that a statistical description is required.

The nuclear level density near the Fermi surface plays an important role in statistical calculation of various fields such as nuclear physics, spallation neutron measurements in Healy Ion Collisions. Therefore a sound knowledge of the fundamental parameters of the level density formula and nucleon effective mass is needed for testing the various theoretical models. The level density function $\rho(U) = dN(U)/dU$ where $N(U)$ is the number of levels with energies close to the excitation energy U. It is possible to describe empirically the function $N(U)$ by adopting the analytical form

$$ln N(U) = U - U_0/T \qquad (1)$$

Equation (1) is known as the temperature law, where $U_0$ and T are adjustable parameters. Thus at low energies the level densities seem to increase exponentially with the energy. But beyond an excitation energy of about 7.5 MeV, some discrete levels cannot be distinguished any longer. Basically the Bethe Fermi gas model, Mughabghab and Dunford (1998) of the level density as given by the equation

$$\rho(U, J) = \frac{2J+1}{24 U^{\left(\frac{5}{4}\right)} 2^{\left(\frac{1}{2}\right)} \sigma^3} e^{2(aU)^{\frac{1}{2}}} e^{-\frac{\left(J+\frac{1}{2}\right)^{\frac{1}{2}}}{2\sigma^2}} \qquad (2)$$

has been successfully used for analysing experimental data. Odd-even effects are incorporated through the equation $\chi = \delta_{N,par} + \delta_{Z,par}$ where $\chi$ is the parameter index, if k is even and 0 if odd. Thus it was established $N(U)$ is inversely proportional to $\chi$.

Essentially $\rho(U, J)$ denote the level density for the nuclear states of spin J, $\sigma$ is the spin dispersion parameter, and 'a' is the level density parameter. The relation between the level density parameter 'a' and the single level density $g(E_f)$ at the Fermi surface $E_f$ if given by

$$a = \frac{\pi^2}{6} g(E_f) \qquad (3)$$

From the Thomas-Fermi model for a finite nucleus, an expansion of the level density in powers of $A^{1/3}$ is given by

$$a^* = A(a_v + a_s B_s A^{-\frac{1}{3}} + a_c B_k A^{-\frac{2}{3}}) \qquad (4)$$

Shell effects can be manifested by incorporating the following relation into equation (4)

$$a = a^* \left[1 + \left(\frac{\delta W}{U}\right)(1 - \exp^{-\gamma U})\right] \qquad (5)$$

where $\quad \gamma = \gamma_0/A^{\frac{1}{3}}$

now $a^*$ is the smooth value of the level density, $a_v, a_s$ and $a_c$ are volume, surface and curvature components of the level density parameter respectively. In equation (5) $\delta W$ is a liquid drop shell correction, $\gamma$ is the shell damping factor which depends on mass numbers. From the knowledge of $\gamma$ and subsequently shell effects there is evidence of the impact on spin dispersion parameter.

From the experimental data on the density of 'n' resonances, we get candid and accurate formation of the level density, Ignatyuk et al. (1975). The presence of deep minima near magic numbers nuclei when nuclear density $\rho$ is studied as a function of mass numbers indicates that shell effects play important role in determining the properties of nuclei. The odd-even effects can be minimized by subtracting from the excitation energy 'U' a quantity $\Delta$ which depends on the parity number $\chi$. Considering that fermions have the tendency to form pairs, and therefore certain energy should be supplied to break these pairs, in addition to the energy needed to excite the fermions. Thus $\Delta$ is the energy required to break the nucleon pairs, so that the final state of the nucleus is analogous to a gas of independent particles. The energy to be given is a constant $(\delta_n + \delta_p)$. Therefore one takes as effective excitation energy of the value

$$E_0 = U - \Delta \qquad (6)$$

Whereby one gets the following equation for the level density

$$\rho(N, Z, U) = \frac{\sqrt{\pi}}{12} \frac{\exp\left(2\sqrt{aE_0}\right)}{a^{\frac{1}{4}}(E_0)^{\frac{5}{4}}} \qquad (7)$$

$\Delta$ is defined as

$\Delta = \delta_n + \delta_p$ for an even-even nucleus

$\Delta = \delta_p$ for odd number of neutrons

$\Delta = \delta_n$ for an odd number of protons

$\Delta = 0$ for an odd-odd nucleus.

A systematic study of mass difference between nuclei reveals that

$$\delta_n = \delta_p = 12/\sqrt{A} \qquad (8)$$

Therefore $\Delta = \chi(12/\sqrt{A})$

Thus, on comparing with experimental data, we see that by introducing, the level density parameter 'a' is less sensitive to even-odd effects. Thus we have a Back-Shifted Fermi gas where only the parameters $\Delta$ and a can be adjusted to reproduce the experimental data for energies of the order of neutron binding energy.

The parameters governing the level density model are presented in Table 1. The discrete level schemes are adopted from the Reference Input Parameters Library (RIPL-2) database. We checked the completeness of the discrete level schemes adopted by correlating the level density and parity ratio of the discrete levels Al-Quraishi et al. (2003). It was observed that except for the $^{58}$Fe level scheme, the level density and parity ratio shows a reasonable behaviour suggesting the completeness of the discrete levels considered in the calculation.

In pre-equilibrium calculations, the two component exciton model using Kalbach systematics, Kalbach (1986) with the particle- hole state density by Dobes and Betak (1983), Betak and Dobes (1976) has been adopted. Their formula is based on the assumption of equidistant level spacing and is corrected for the effect of the Pauli Exclusion Principle and for the finite depth of the potential well has been incorporated. Fu's pairing correction Fu (1984) has also been taken into account. For the single-particle state densities, $g_\pi = Z/15$ and $g_Y$= N/15.   The theoretical calculations are compared with the experimental data from EXFOR database.

## 3. Results and discussions

The excitation curve for $^{54}$Fe(n, p)$^{54}$Mn is given in Fig. 1.

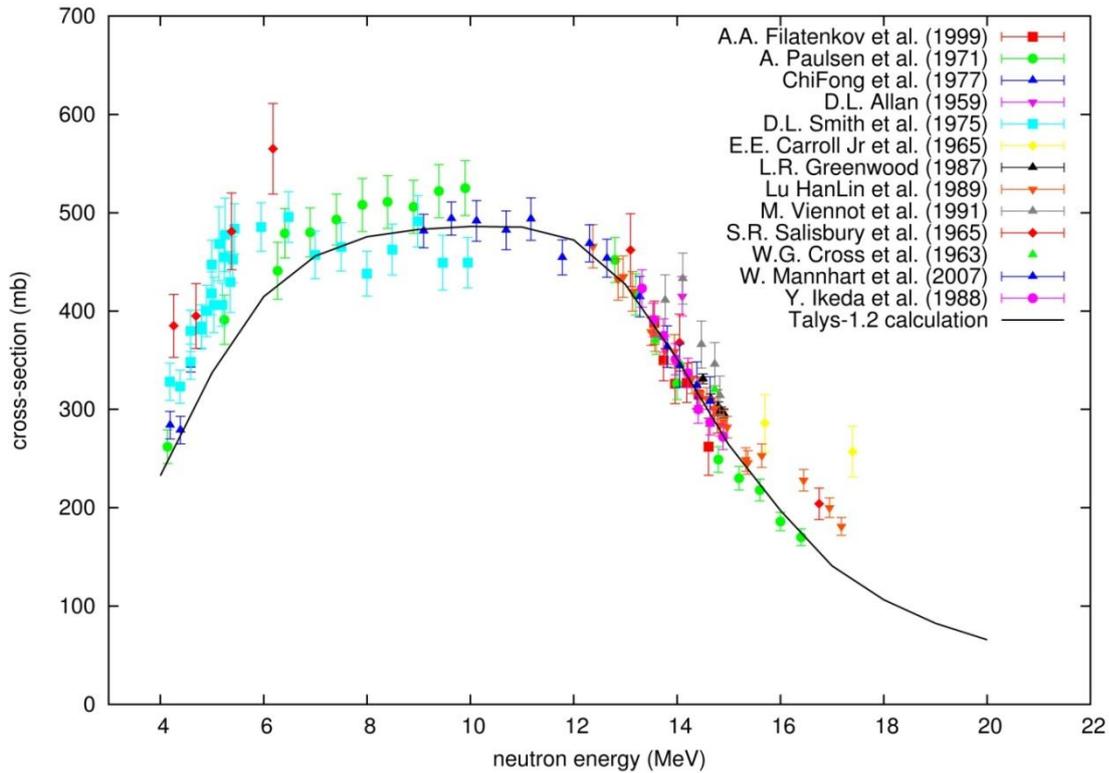

Fig. 1.

The investigation using Talys-1.2 indicates that the emitted proton spectra, the compound nucleus contribution is dominated by the low energy protons and the pre-equilibrium contribution mainly comes from the high energy emitted protons which is also reflected in the total (n, p) cross-section.  In Fig. 1, the bell like shape of the excitation curve is a typical characteristic of compound nucleus which rises abruptly above the reaction threshold and descends due to the competitive (n, np) reaction and increase of the pre-equilibrium contribution with neutron energy.  The lower energy part is dominated by compound nucleus contribution and the pre-equilibrium tail which appears around 16-18 MeV.

The excitation curve for $^{56}$Fe (n, p)$^{56Mn}$ is given in Fig. 2.

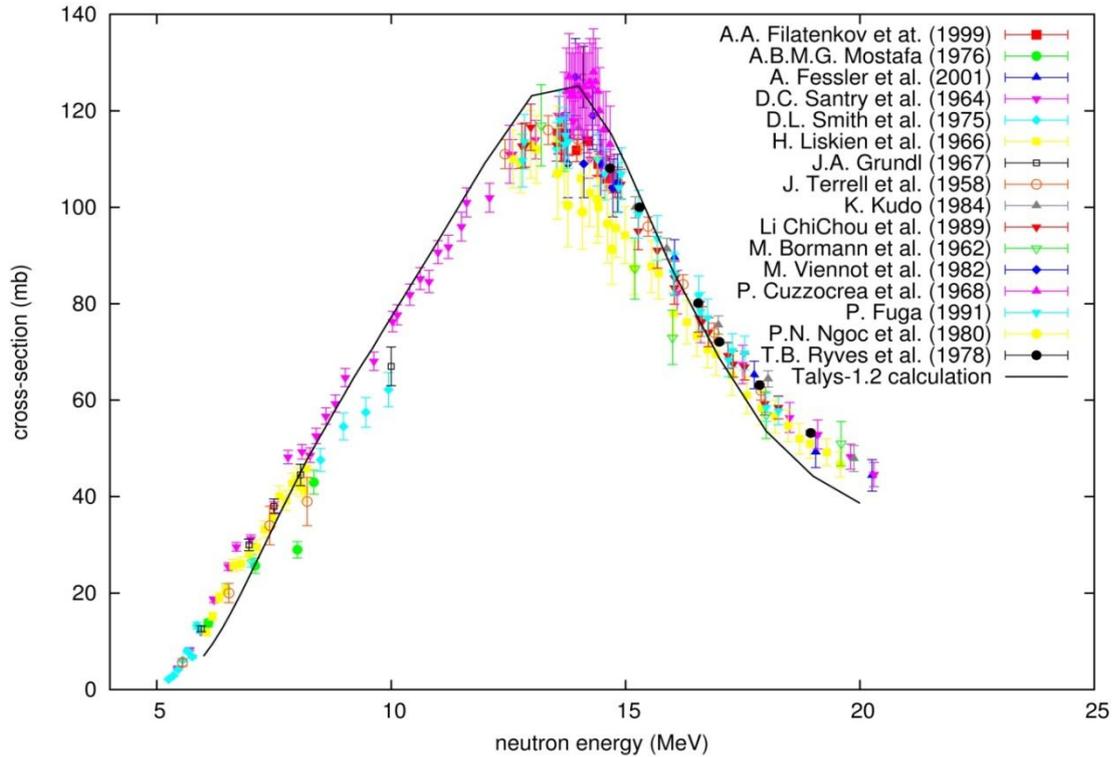

Fig. 2.

All the parameters are calculated in a similar manner as done for the previous reaction except the shell damping factor. The overall agreement between the measured and calculated cross-section is excellent. By comparing the Fig. 1. and Fig. 2, the cross-section rises slowly with neutron energy in Fig. 2. than Fig. 1. indicating the increasing contribution of pre-equilibrium with the increase of neutron number.

The excitation curve for $^{57}Fe(n, p)^{57}Mn$ is given in Fig. 3. This is the only odd-A target which is studied here. The disappearance of the bell like shape indicating the increase contribution of pre-equilibrium which is an agreement with the isotopic effect.

The excitation curve for $^{58}Fe(n, p)^{58}Mn$ is given in Fig. 4. There exists only few experimental data around 14-15 MeV and with a lot of discrepancies. The data from ENDF/B-VII is also plotted along with the present model calculation. The free parameter $\gamma_0$ turns out to be zero which essentially implies that the shell damping factor is not important.

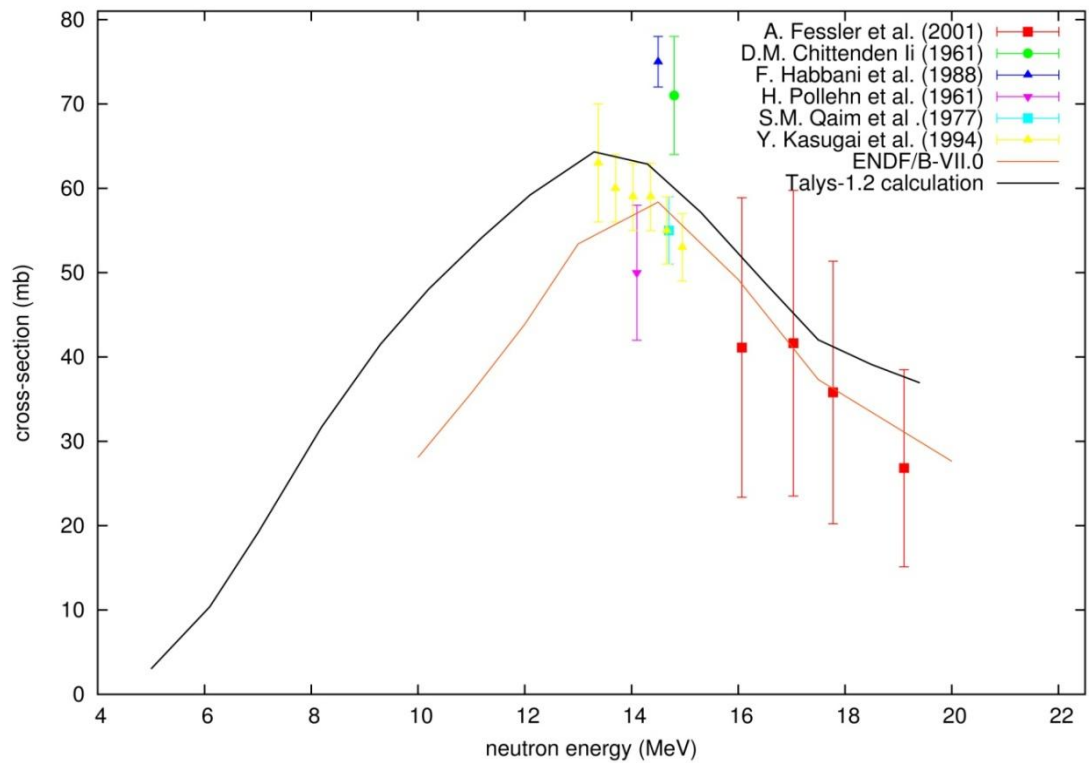

Fig. 3.

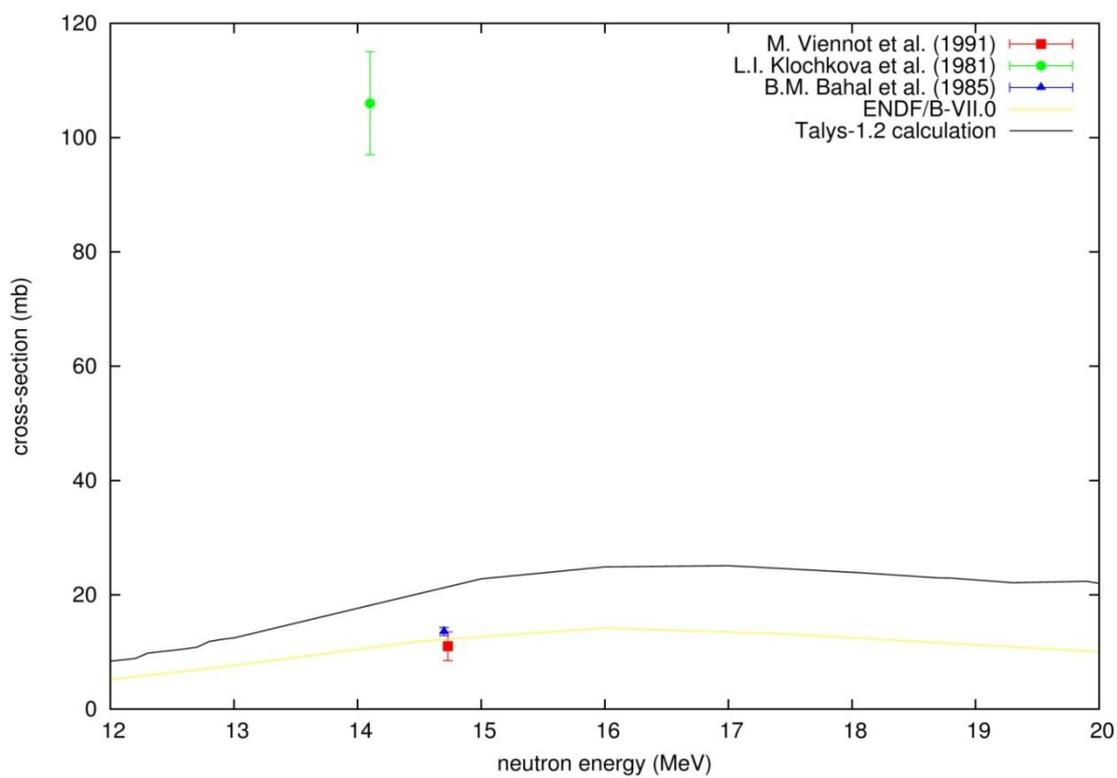

Fig. 4.

The excitation curve for the (n, p) reaction in the $^{54}$Fe behaves in a way typical for a compound nucleus decay. The high energy pre-equilibrium tail appears only at neutron energies above 16 MeV. On the other hand, for all heavier Iron isotopes the cross-sections are low and their monotonic increase with neutron energy simulates the high energy part of the excitation curve observed for the $^{54}$Fe target. Such behaviours of the cross-section indicate a pre-equilibrium emission. The observation from all the excitation curves suggested that the pre-equilibrium contribution to the total (n, p) cross-section increases with the mass number. Fig. 5 is a plot for cross-section versus atomic mass number for Fe isotopes at the projectile energy of 14.7 MeV. It is observed from the Fig. 5 the decrease in the (n,p) reaction cross-section with increasing mass number A of the isotope.

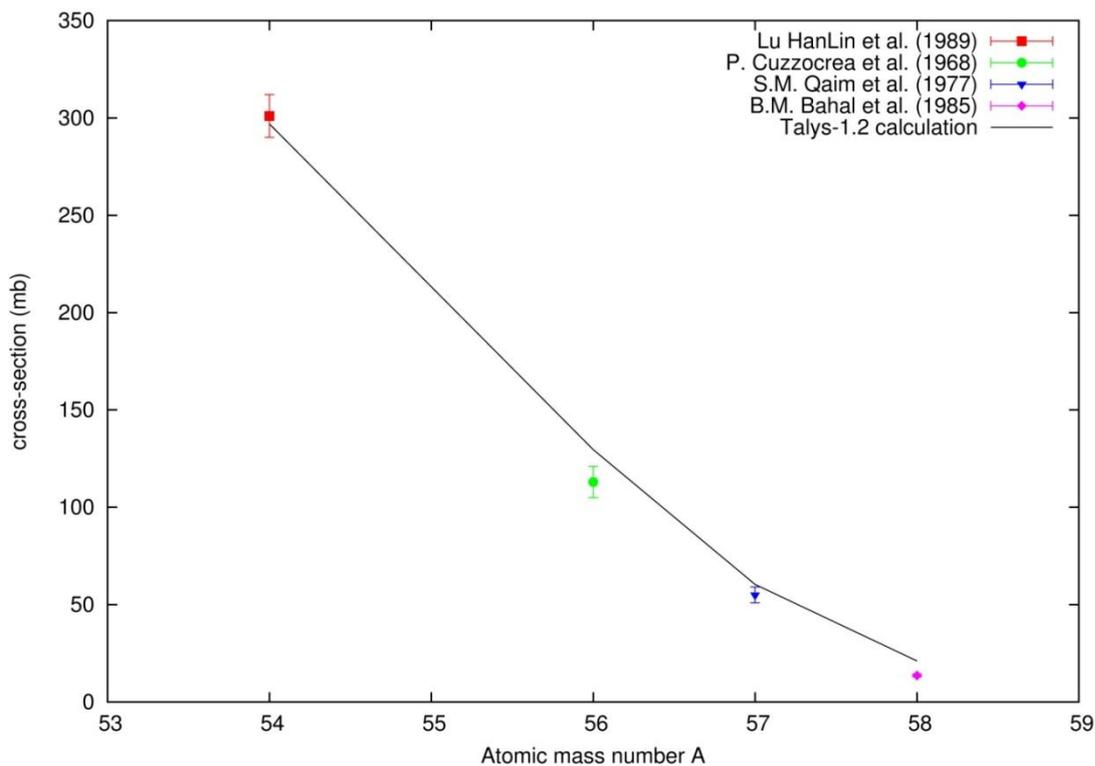

Fig. 5.

This was pointed out by Gardner (1962) as a Q-value effect. Q-value versus the asymmetry parameter is given in Fig. 6. The systematic decrease in (n, p) cross-sections with increasing neutron number in reactions induced by neutrons on isotopes of Iron can be explained in terms of the pre-equilibrium model.

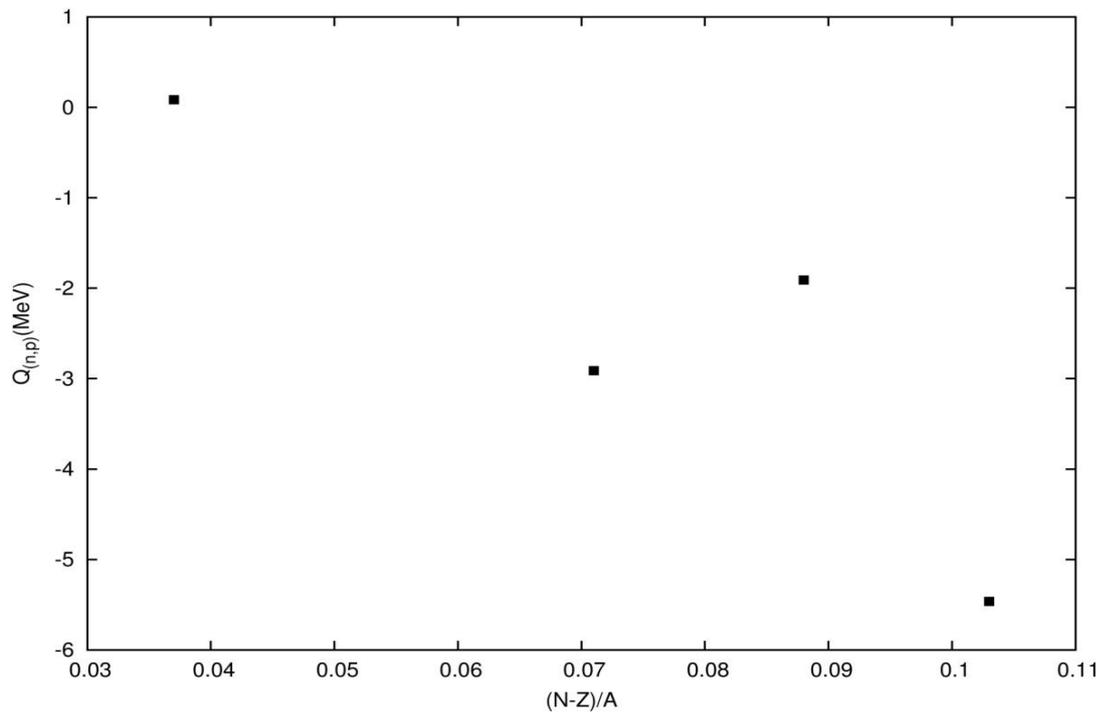

Fig. 6.

| Nucleus | $a$ $MeV^{-1}$ | $a^*$ $MeV^{-1}$ | $\gamma$ $MeV^{-1}$ | $P$ $MeV$ | $E_{sh}$ $MeV$ | $T$ $MeV$ | $E_0$ $MeV$ |
|---|---|---|---|---|---|---|---|
| 54Fe | 5.574 | 7.292 | 0.1214 | 3.265 | -3.368 | 1.729 | -1.472 |
| 56Fe | 6.269 | 7.516 | 0.1199 | 3.207 | -2.150 | 1.683 | -2.697 |
| 57Fe | 6.701 | 7.627 | 0.1360 | 1.589 | -1.310 | 1.407 | -1.740 |
| 58Fe | 7.739 | 7.739 | 0. | 3.151 | -0.528 | 1.321 | -0.461 |

Table 1.

## 4. Conclusions

The level density parameters for the various nuclides have been determined by several experimental methods in neutron, charged particle studies as well as from measurements of medium-energy HIC. Some of these methods also include average spacing's of n resonances, inelastic n scattering, evaporation spectra and capture measurements. We have analyzed the excitation functions of stable isotopes of iron which are important construction material components in advanced fission and fusion reactions with recent code Talys-1.2. We get overall good agreement between the measured data and model calculated values have been obtained using Talys-1.2 by varying only one free parameter called shell damping parameter. The (n, p) cross-section is very sensitive to a slight change in the value of shell damping parameter choices. We observed that a better description on the dependence of shell damping parameter on mass number and choices on $\gamma_0$ would evidently lead to a greater predictive power of the nuclear model. We have also observed that the (n, p) cross-section for stable Fe isotopes decrease with the increase in the neutron number and the pre-equilibrium contribution increases with the increase in the neutron number for (n, p) reaction in Fe isotopes.

**Acknowledgement**

The authors (Jyrwa and Joseph) are extremely grateful to Dr. S. Ganesan RPDD, BARC Trombay, for his valuable suggestions and ideas, which was based on similar lines from the work of Lalremruata et al. (2009) Published in Annals of Nuclear Energy.